\def\n{\noindent }
\def\mbf#1{{\mathbf {#1}}}
\def\eq{\ =\ }
\def\mns{\ -\ }
\def\pls{\ +\ }
\def\be{\begin{equation}}
\def\ee{\end{equation}}

\documentclass[12pt,a4paper,epsfig]{iopart}
\usepackage{epsfig}
\begin{document}
\title[Electronic and Optical Properties of ZnIn$_2$Te$_4$]
      {Electronic and Optical Properties of ZnIn$_2$Te$_4$}
\author{\bf Biplab Ganguli, Kamal Krishna Saha}
\author{\bf Tanusri Saha-Dasgupta and Abhijit Mookerjee}
\address{S. N. Bose National Centre for Basic Sciences.
Block-JD, Sector-III, Kolkata-700098, India.}
\author{\bf A.K.Bhattacharya}
\address{Centre for Catalysis and Materials Design, Department of Engineering, 
University of Warwick, Coventry, U.K.}
\begin{abstract} Band structure and optical properties of defect- 
Chalcopyrite type semiconductor ZnIn$_2$Te$_4$ have been studied by TB-LMTO
 first principle technique. The optical absorption calculation suggest that 
ZnIn$_2$Te$_4$ is a direct-gap semiconductor having a band gap of 1.40 eV., 
which confirms the experimentally measured value. The calculated complex 
dielectric-function $\epsilon(E) = \epsilon_1(E) + i\epsilon_2(E)$ 
reveal distinct structures at energies of the critical points in the 
Brillouin zone.
\end{abstract} 
\jl{3}
\ead{biplabg@bose.res.in}
\pacs{71.20,71,20c}
\maketitle
\section{Introduction}
The ternary semiconducting compounds $A^{II}B_2^{III}C_4^{VI}$ have been widely 
investigated because of their potential applications in electro-optic, 
optoelectronic, and non-linear optical devices \cite{Georg}. 
Most of these compounds have the
defect chalcopyrite or  stanite structure \cite{Hahn,Madel}. 
In an $A^{II}B_2^{III}C_4^{VI}$ defect chalcopyrite compound, 
A, B, and C atoms and the vacancy E, are distributed as 
follows \cite{Madel},: A site at (a/2,0.c/4), B at (0,0,c/2) and (0,a/2,c/4), C at 
$(\alpha,\overline\beta,\gamma)$, $(\overline\alpha,\beta,\gamma)$, 
$(\beta,\alpha,\overline\gamma)$ and $(\overline\beta,\overline\alpha,
\overline\gamma)$ and E at (0,0,0). It's lattice parameters are given by 
$\alpha = \beta = a/4, \gamma = c/8$ and c = 2a, as shown in the figure~1.

In this communication we shall investigate the electronic structure and optical 
properties of the defect chalcopyrite $ZnIn_2Te_4$ from a first principles approach. 
The chosen system is not only important for 
its application in technology, but is also important for the study of the effect 
of defect or impurity on various properties of materials. There is also growing 
interest in the study of defect and nano structures \cite{Myers,Domain}. 
One of the experimental method for the preparation of nano-structures is to fill 
in the structural voids in a ``host" compound with atoms of a given substance. 
These structures could either be local void clusters or even tubulary voids. 
The characteristics dimensions. The host compound helps to form a matrix in which 
these nano structures stabilize. In a subsequent communication, we shall extend 
the present work to include nano structures in host materials.

\begin{figure}[b]
\centering
\epsfxsize=5in \epsfysize=5in
\rotatebox{0}{\epsfbox{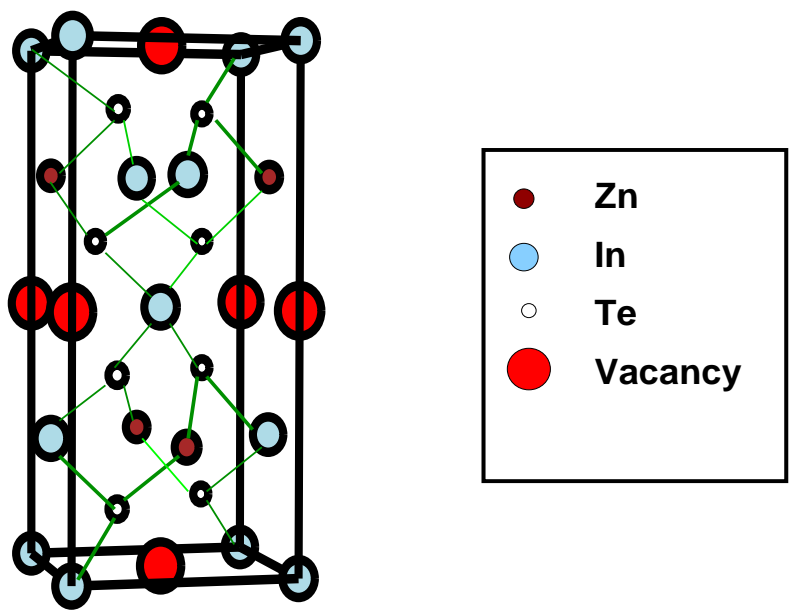}}
\vskip -2cm
\caption{Crystal structure of defect-chalcopyrite-type semiconductor
$ZnIn_2Te_4$. Here an orthorhombic  primitive cell is shown }
\end{figure}

Recently Ozaki \etal \cite{Matsu,Ozaki} have carried out a detailed 
experimental and theoretical study of optical properties of amorphous 
and crystalline $ZnIn_2Te_4$. There are few other experimental  
measurement  \cite{Neumann}- \cite{Manca1} on the compound also 
but there has been no
first principles calculation  for the electronic structure and optical 
properties of this material.

The band structure calculations by empirical parameterized tight-binding methods 
has been carried out for  $ZnIn_2Te_4$ by Meloni 
\etal  \cite{Meloni} and Ozaki \etal  \cite {Ozaki}. These calculations  
required fitted parameters. The actual
crystal structures also seem to have been simplified. Meloni \etal have assumed a 
pseudocubic structure (space group = $V_d$), rather than the actual chalcopyrite 
type structure. Ozaki \etal have taken a defect chalcopyrite structure. However, the  
space group $S_4^2$, which they assumed, does not reflect the correct 
symmetry of this structure. 
This is the usual chalcopyrite structure but with vacancies
at the sites as shown in the figure 1.  Because of the vacancies, 
this structure does  not have the full space group $S^2_4$. 
Rather, we can assign this structure the $I\overline4$
symmetry. The positions of Te's are zinc-blende type, {\sl i.e.} two inter-penetrating fcc
lattices shifted one-fourth of the way along a body diagonal. The values of the lattice parameters $a$ and $c$are taken
from experiment to be 6.11$\AA$  and 12.22$\AA$ respectively as reported  by Hahn and Ozaki
 \cite{Hahn,Ozaki}.
The positions of the atoms within the unit cell is shown in table 1.

\begin{table}[t]
\centering
\begin{tabular}{|c|ccc|| c|ccc|}\hline
Atom type & \multicolumn{3}{c||}{Positions} & Atom type & \multicolumn{3}{c|}{Positions}\\ \hline
 Zn   & 0.00  & 0.50 & -0.50 &
 In1   & 0.00  & 0.00  & 0.00\\
 In2   & 0.50  & 0.00 & -0.50 &
 Te   & 0.25  & 0.25  & 0.25\\
 Te   & 0.25 & -0.25 & -0.25 &
 Te  & -0.25  & 0.25 & -0.25\\
 Te   & 0.25  & 0.25 &-0.75 &
 E1    & 0.00  & 0.00 & 1.00\\
 E2    & 0.25  & 0.25 &-0.25 &
 E2    & 0.25 & -0.25  & 0.25\\
 E2   & -0.25  & 0.25  & 0.25 &
 E2  &   0.25  & 0.25  &-1.25\\
 E3  &   0.50  & 0.00  & 0.00 &
 E3  &   0.00  & 0.50  & 0.00\\
 E4  &   0.00  & 0.00  & 0.50 &
 E4  &   0.00  & 0.00 & -0.50 \\  \hline
\end{tabular}
\caption{Positions within the unit cell of atomic basis, including the
empty spheres to take into account the voids within the structure}
\end{table}

\section{Results and Discussion}

\subsection{Electronic Structure Calculations}

To start with, for our electronic structure calculations we have used 
the well established 
tight-binding linearized muffin-tin orbitals method (TB-LMTO), discussed 
in detail elsewhere \cite{Ander1,Ander2,Skriver}. Electron correlations are
taken  within local density approximation 
of density functional theory \cite{Kohn1,Kohn2}.

\begin{figure}
\centering
\epsfxsize=5in \epsfysize=4.5in \rotatebox{270}{\epsfbox{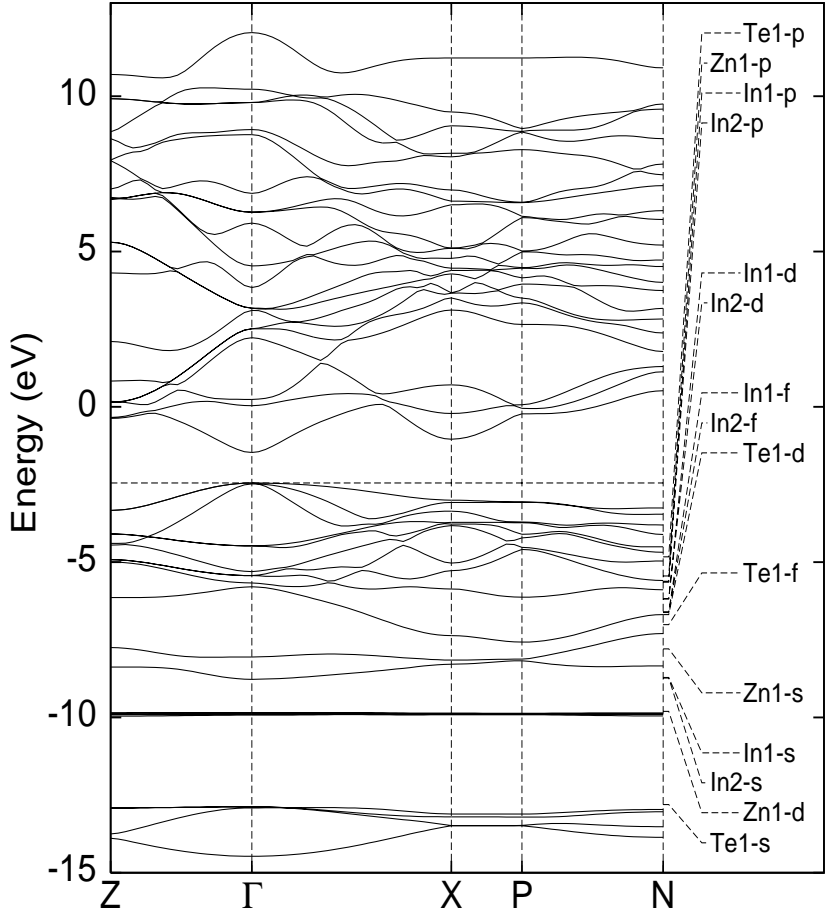}}
\caption{The band structure of $ZnIn_2Te_4$ within the TB-LMTO}
\label{bnd1}
\end{figure}

The TB-LMTO energy bands in several high symmetry directions in reciprocal space is shown in
figure \ref{bnd1}.  Let us examine this in some detail. 
The basis of the
TB-LMTO starts from the minimal set of muffin-tin orbitals of a KKR formalism and then linearizes it by expanding
around a `nodal' energy point $E_{\nu \ell}^\alpha$. The wave-function is then expanded
 in this basis :                

\be
\Phi_{j\mbf{k}}(\mbf{r})  \eq  \sum_{L}\sum_{\alpha} c^{j\mbf{k}}_{L\alpha}\left[ \phi_{\nu L}^\alpha
(\mbf{r})+\sum_{L'}\sum_{\alpha '}\ h_{LL'}^{\alpha\alpha '}(\mbf{k})\ {\mathaccent 95 \phi}_{\nu L'}
^{\alpha '}(\mbf{r})\right]\nonumber
\ee

\noindent and,

\begin{eqnarray*}
\phi_{\nu L}^\alpha({\mbf{r}}) & \eq & \imath^{\ell}\ Y_L(\hat{r})\ \phi_\ell^\alpha(r,E_{\nu\ell}^\alpha) \\
{\mathaccent 95 \phi}_{\nu L}^\alpha(\mbf{r}) & \eq & \rule{0mm}{6mm}\imath^{\ell}\ Y_L(\hat{r})\ \frac{\partial
\phi_\ell^\alpha(r,E_{\nu \ell}^\alpha)}{\partial E} \\
h_{LL'}^{\alpha\alpha '}(\mbf{k}) & \eq & (C_L^\alpha - E_{\nu \ell}^\alpha)\ \delta_{LL'}\delta_{\alpha\alpha '} \pls \sqrt{\Delta_{L}^\alpha}\ S_{LL'}^{\alpha\alpha '}(\mbf{k})\ \sqrt{\Delta_{L'}^{\alpha '}} \\
\end{eqnarray*}
 \noindent$C_L^\alpha$ and $\Delta_L^\alpha$ are TB-LMTO potential parameters and  $S_{LL'}^{\alpha\alpha '}(\mbf{k})$ is the structure matrix. 

First note that for the calculation of the optical properties of the solid we need to span a large
energy range, from the occupied valence to the unoccupied conduction states. For ZnInTe this spans
a range from -15 eV to 10 eV. The ``nodal" energies cluster below -5 eV in the valence band. Obviously,
the conduction part of the band is not accurately reproduced. 
The third generation NMTO expands the wavefunction in term of a basis which is expanded as a Lagrange interpolation around
a discrete set of {\sl nodal} energies $\{\epsilon_n\}$ :

\be
\Phi_{j\mbf{k}}(\mbf{r}) \ =\   \sum_{L}\sum_{\alpha}\ c^{j\mbf{k}}_{L\alpha}\ \left[\rule{0mm}{4mm}
 \sum_{n=0}^N\ \phi_{nL}^{\alpha}(\mbf{r})\ {\cal L}^{(N)}_{n,LL'}(\mbf{k})\right]
\ee

\n The ${\cal L}^{(N)}_{nRL,R'L'}$ are Lagrange matrices which are such that the energy dependent partial wave basis $\phi_{L}^\alpha(\mbf{r},E)$  takes the values $\phi_{nL}^{\alpha}(\mbf{r})$ at the {\sl nodal} energies.
Unlike the LMTO, the {\sl nodal} energies are independent of the indeces $L\alpha$. By choosing the {\sl nodal} energies across the energy range of interest we can accurately reproduce the bands in that range. This was shown for GaAs in the range -15~eV to 20~eV by
 Andersen and Saha-Dasgupta  \cite{Ander3}.

\begin{figure}
\centering
\epsfxsize=5in \epsfysize=4.5in \rotatebox{270}{\epsfbox{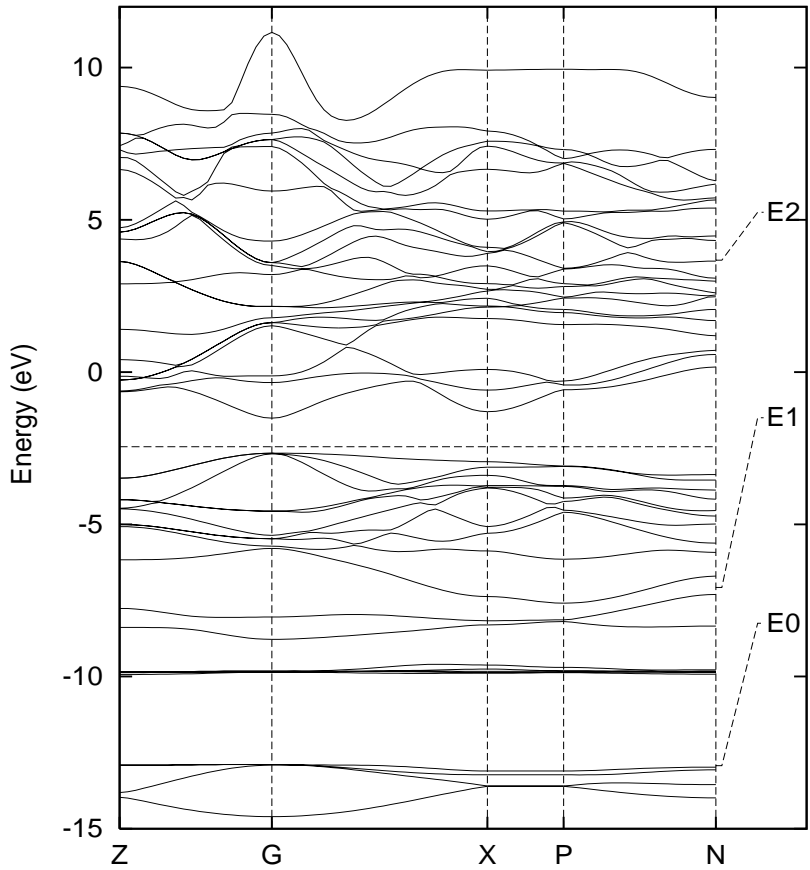}}
\caption{The band structure of $ZnIn_2Te_4$ within the NMTO}
\label{bnd2}
\end{figure}

The figure \ref{bnd2} shows the energy bands obtained from the NMTO using three {\sl nodal} energies
spread across the energy range of interest. In comparison with the figure \ref{bnd1}, we note that
the largest change occurs in the conduction band, away from the TB-LMTO nodal energies bunched below 
-5 eV. 
We have used the von Barth-Hedin exchange \cite{Barth} with 512 {\bf k}-points in the irreducible part of the
Brillouin zone. 

\begin{figure}
\centering
\epsfxsize=3in \epsfysize=5in
\rotatebox{-90}{\epsfbox{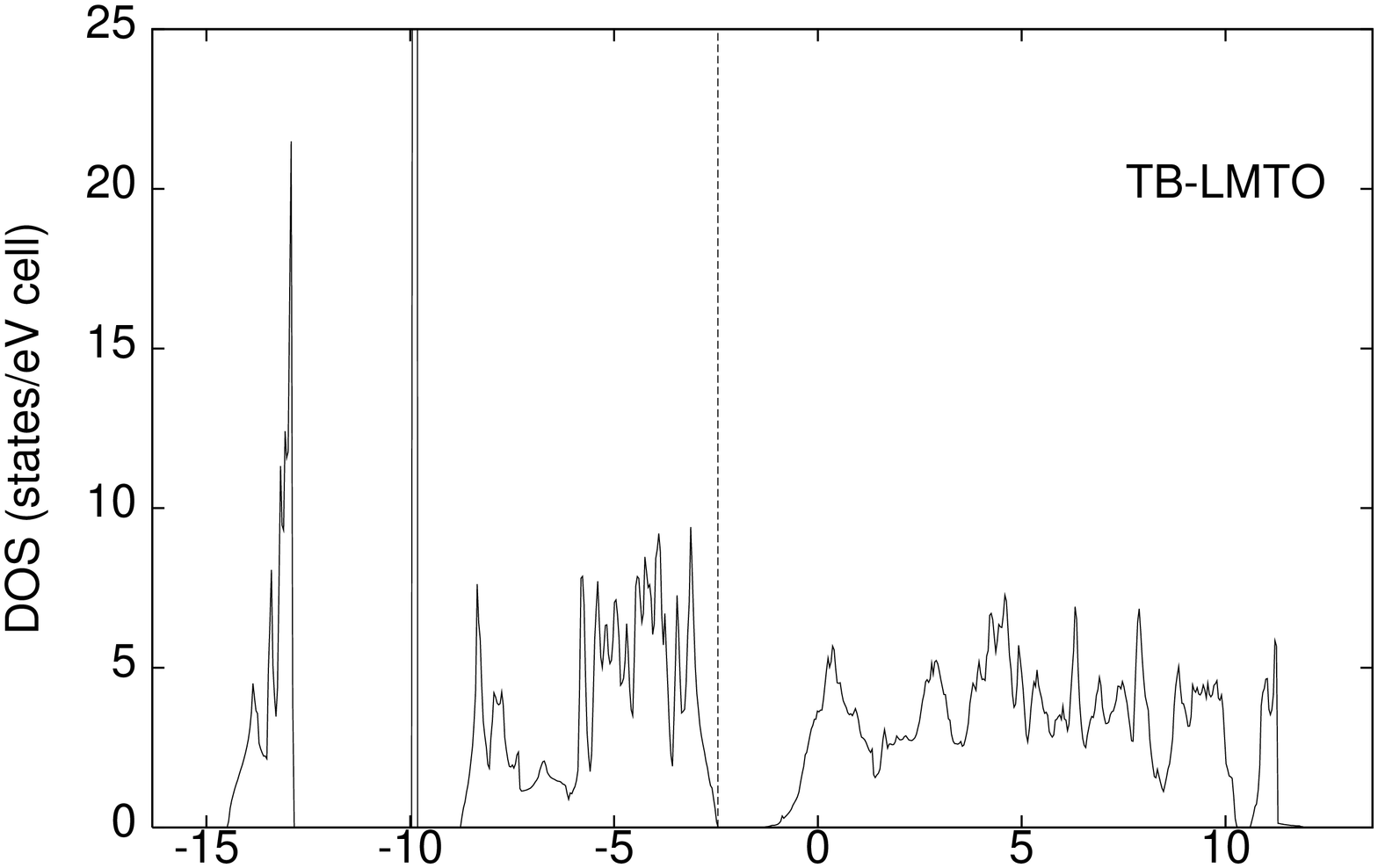}}
\epsfxsize=3in \epsfysize=5in
\rotatebox{-90}{\epsfbox{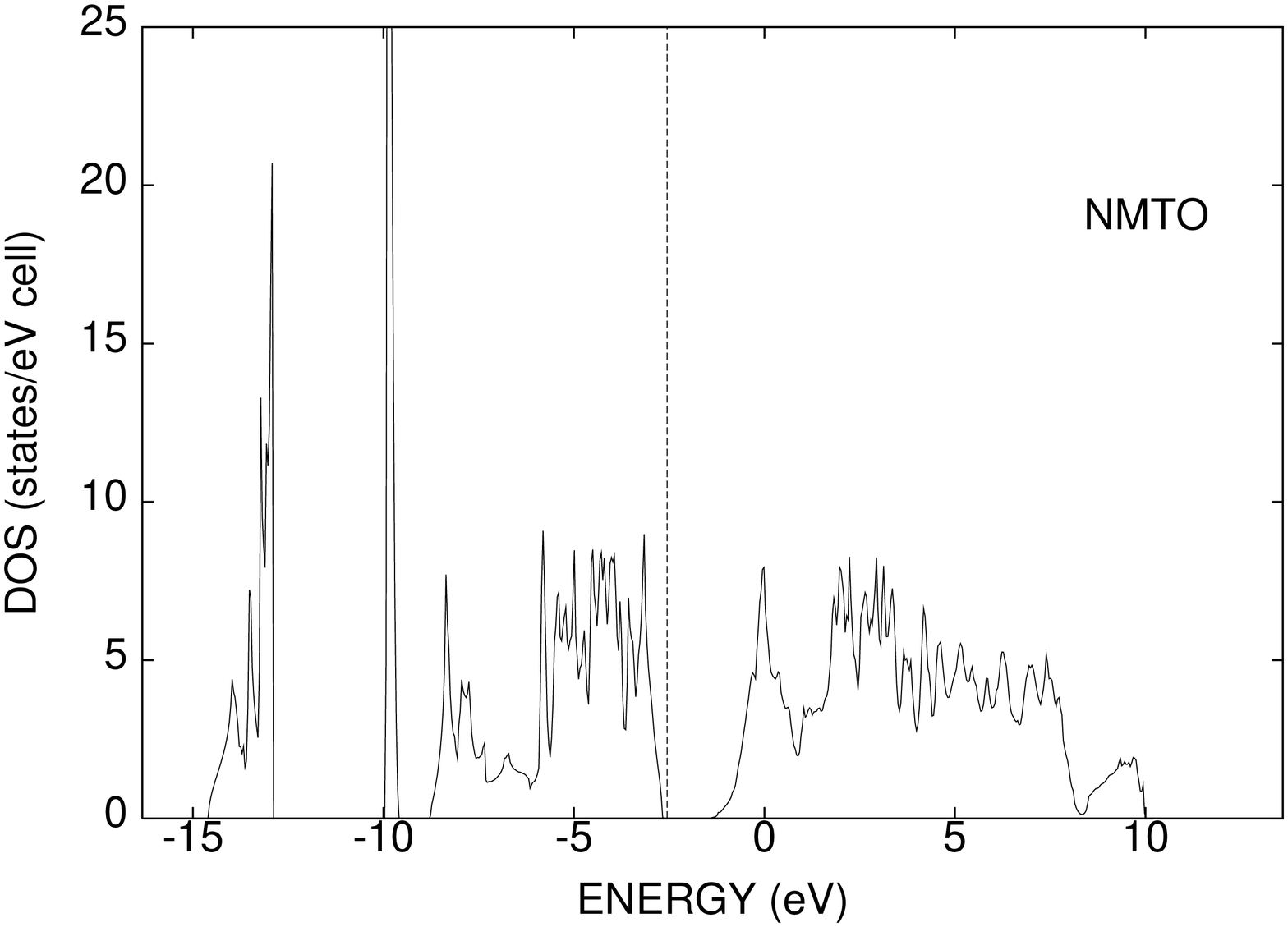}}
\vskip 0.5cm
\caption{Density of States for $ZnIn_2Te_4$ (top) calculated from the TB-LMTO and (bottom) calculated from NMTO
with three nodal energies across the energy range}
\label{dos}
\end{figure}

\begin{figure}[t]
\centering
\epsfxsize=5in \epsfysize=5in
\rotatebox{0}{\epsfbox{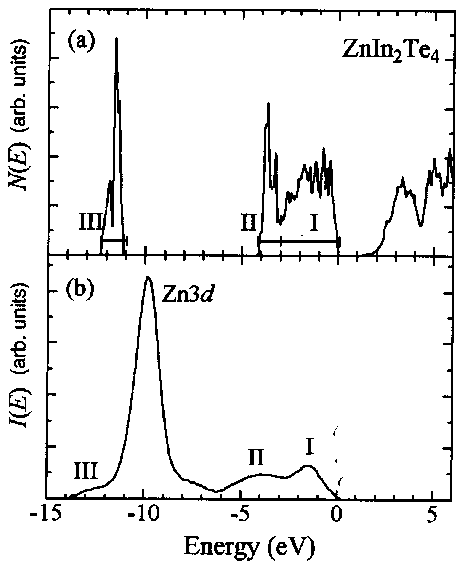}}
\caption{(a)Theoretical density of states N(E) and (b) experimental XPS spectrum I(E) taken from 
            Ref. \cite{Ozaki}}
\label{xps}
\end{figure}

The calculated density of states  within TB-LMTO and NMTO are shown in the figure \ref{dos}.
 It is evident from these two figures that there is  significant difference between 
these two only in the conduction band. From the band structure comparison, this was also
evident for the reasons discussed earlier.  
These densities  agree quite well with the 
experimental XPS measurements of Ozaki \etal \cite{Ozaki} and also with their parametrized tight-binding
calculations. These results are shown in the 
figure \ref{xps}. 
It must be noted here that since these calculations are based on the LDA, we do obtain a lower band gap.
In this case the NMTO estimate of the band gap is around 1.3 eV, while the XPS data indicate a
gap of around 2 eV.  Nor can we have much confidence about the conduction bands. 
This problem can be tackled more accurately, for example,  
through by quasi-particle band structure within a GW  type approximation \cite{ssp54}. 
These NMTO based calculations are a much better starting point of GW self-consistency iterations
than the first or second generation LMTOs (see comments in \cite{Ander3})

The TB-LMTO band structure gives us insight into the various structures in the density of states. 
At the lowest energies around -14 eV we have the states arising out of the Te-$s$ electrons. The
sharp peak at -10 eV is due to the very narrow Zn-$d$ bands. The next structures around -8 eV
arise from the In-$s$ and Zn-$s$ electrons. The predominantly covalently bonded Te $sp$ states
gives rise to bonding and anti-bonding bands around -5 eV and 5 eV respectively. The band gap lies 
between these bands.

\subsection{Optical Properties:}

In recent years a number of methods have been proposed for calculating optical properties 
within the framework of the LMTO  \cite{Us,Al,Ze0,Ho} for both metals \cite{Us,Al} and 
semi-conductors \cite{Al2,Ze,Ho}. 
Upenski \etal \cite{Us} proposed a method for the accurate calculation of the optical matrix 
elements based on the continuity equation for the charge -density operator. They were 
able to calculate accurate optical matrix elements by including the combined correction 
term to compensate for inaccuracies in the wavefunction due to the basis set in LMTO 
theory being finite. This method and corrections were subsequently applied by other authors
 \cite{Al,Al2} to calculate accurate optical spectra of Ge, GaAs, InSb and CdTe.

In all the above calculations however, the gradient operator has been employed for the 
determination of the optical matrix. There is an alternative method developed by 
Hobbs \etal \cite{Ho} which avoids the determination of gradient operator. Their method 
allows for the inclusion of non-local potentials in the Hamiltonian. In their method 
they employed Green's second identity and the commutation relation between the position 
and Hamiltonian operators. They finally wrote the momentum matrices in terms of 
Gaunt coefficients \cite{Rose} and potential parameters which are defined within the LMTO 
method \cite{Ander1,Skriver}. The mathematics has been described in detail both by Hobbs
\etal\ and by Saha \etal (\cite{Saha}). We shall indicate here the minor changes required
for the calculation of the transition matrix element within the NMTO.

The expression for the imaginary part of the dielectric response remains the same, as
derived from the Kubo formula :

\be
\fl \epsilon_{2}^\gamma(\omega)\eq \frac{-8{\pi}^{2} e^{2}}{ 2m^{*^{2}} \Omega} \frac{1}{{\omega}^2}
\sum_{i}\sum_{f} {\vert \langle \psi_{f} \vert {\mathaccent 94 e}
_{\gamma}\cdot{\mathaccent 94 p} \vert \psi_{i}\rangle \vert}^{2}F_{i}(1-F_{f})
\delta (E_{f}-E_{i}-\hbar \omega )
\ee

\n where, 
$ m^{*}$ is the effective mass of the electron,$ \Omega $ is the volume of the
sample, i and f refer to the initial and final states respectively, $ \gamma
$ refers to the direction of polarization of the incoming photon, $
{\mathaccent 94 p}$ is the momentum of the electron and $ F_{i,f}$ is the
occupation probability of the initial and final states respectively.
For semiconductors, i lies in the valence band and f in the conduction band
and at $0^{0 }K$ we have $F_{i} = 1$ and $F_{f} = 0$.

We can obtain the real part of the dielectric function $\epsilon_1(\omega)$ from a Kramers-Kr\"onig
relationship.

\begin{equation}
\epsilon_{1}(\omega)\eq 1 \pls \frac{2}{\pi}\int_{0}^{\infty} \frac{(\omega^{\prime}\mns\omega)
\ \epsilon_{2}(\omega^{\prime})}{\omega^{\prime^2}\mns\omega{^2}}d\omega^{\prime}
\end{equation}

The transition probability which involves momentum matrix elements can be evaluated 
using a gauge independent formalism and commutation relation
\vskip 1cm
\begin{eqnarray}
\mbf{P} \eq m_e \  \frac{d\mbf{r}}{dt}\nonumber 
\eq & \frac{m_e}{\imath\hbar} \left[\rule{0mm}{3mm} \mbf{r}, H\right]\nonumber
\end{eqnarray}
\vskip 1cm
Thus calculation of momentum matrix elements means calculation of the  
following integrals:
\vskip 1cm
\begin{equation*}
\int {\phi^{\alpha}_{nL'}(\mbf{r})}^{\star}\ \mbf{r}H\ \phi_{nL}^\alpha (\mbf{r})\ d^3{\mbf{r}}
\hskip .5cm \& \hskip .5cm   
 \int {\phi^{\alpha}_{nL'}(\mbf{r})}^{\star}\ H\ \mbf{r}\ \phi_{nL}^\alpha (\mbf{r})\ d^3\mbf{r}
\end{equation*}
\vskip 1cm
By noting the relations:
\[
\rule{0mm}{6mm} H\ \phi_{nL}^\alpha(\mbf{r})  \eq   \epsilon_{n}^\alpha\ \phi_{n L}^\alpha(\mbf{r})
\]

\n and using Green's second identity we may write the integrals as:

\[
\fl \int {\phi^{\alpha}_{n L'}(\mbf{r})}^{\star}\ \mbf{r}\ H\ \phi_{n L}^\alpha (\mbf{r})\ d^3{\mbf{r}} \eq  \imath^{\ell -\ell '}\ \epsilon_{n}^\alpha\ \mbf{\Gamma}_{LL'}\ \int_{0}^{s_\alpha}
\phi_{n\ell '}^{\alpha}(r)\ \phi_{n\ell}^\alpha (r)\  r^3\ dr\]

\begin{eqnarray*}
\fl \int {\phi^{\alpha}_{n L'}(\mbf{r})}^{\star}\ H\ \mbf{r}\ \phi_{n L}^\alpha (\mbf{r
})\ d^3\mbf{r}   \eq
\imath^{\ell -\ell '} \mbf{\Gamma}_{LL'}\left\{ \epsilon_{n}^\alpha \int_{0}^{s_\alpha} \phi_{n\ell '}^\alpha (r) \ \phi_{n\ell}^\alpha (r) r^3 dr
 \ldots \right.\nonumber \\
\rule{0mm}{6mm} \phantom{xxxxxxxxxx}\ldots  \left.  \pls (\hbar^2/2m_e)s^2_\alpha \ \phi_{n\ell}^\alpha (s_\alpha)\ \phi_{n\ell '}^\alpha (s_\alpha)\        \left( D_{n \ell '}^\alpha - D_{n \ell}^\alpha -1\right)\right\} \nonumber \\
\end{eqnarray*} 
\vskip 1cm
\n where, $s_\alpha$ is the atomic sphere radius of the $\alpha$-th atom in the unit cell
and ${\mbf{\Gamma}}_{LL'}$ is a combination of Gaunt coefficients  \cite{Ho} :

\begin{equation*}
\fl {\mbf{\Gamma}}_{LL'}  \eq   \sqrt{(2\pi/3)}\left[\ \left(G^{m',-1,m}_{\ell ',1,\ell} -  G^{m',1,m}_{\ell ',1,\ell}\right)\ \hat{\mbf{i}} \pls
  \imath \left( G^{m',-1,m}_{\ell ',1,\ell} +  G^{m',1,m}_{\ell ',1,\ell}\right)\ \hat{\mbf{j}} \pls
  \sqrt{2}\ G^{m',0,m}_{\ell ',1,\ell} \ \hat{\mbf{k}}\ \right]
\end{equation*}

\n $D_{n\ell}^\alpha$ is the logarithmic derivative of $\phi_{nu\ell}^\alpha (r)$ at  $r\eq s_\alpha$

\begin{figure}
\centering
\epsfxsize=2.5in \epsfysize=3in
\rotatebox{-90}{\epsfbox{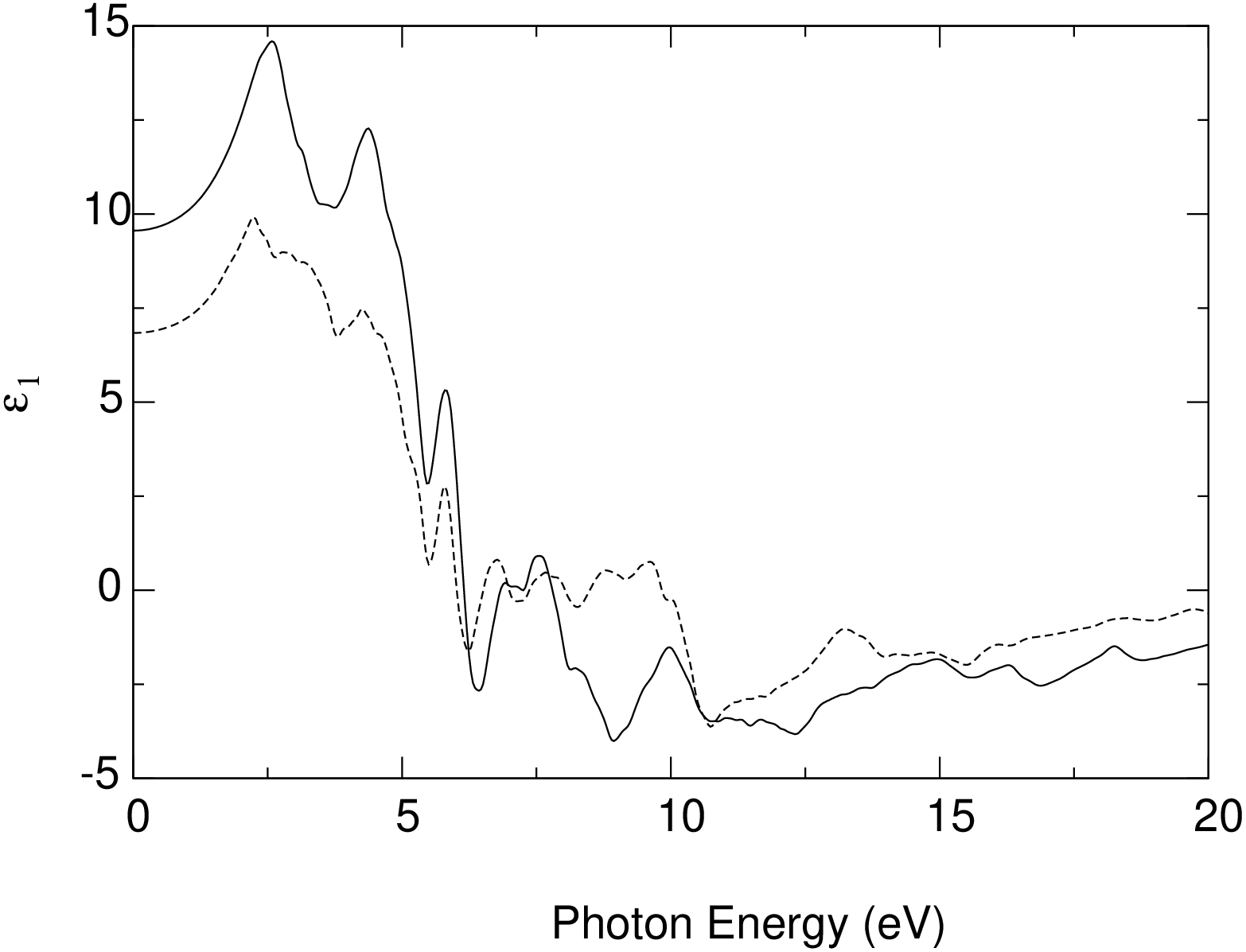}}
\rotatebox{-90}{\epsfbox{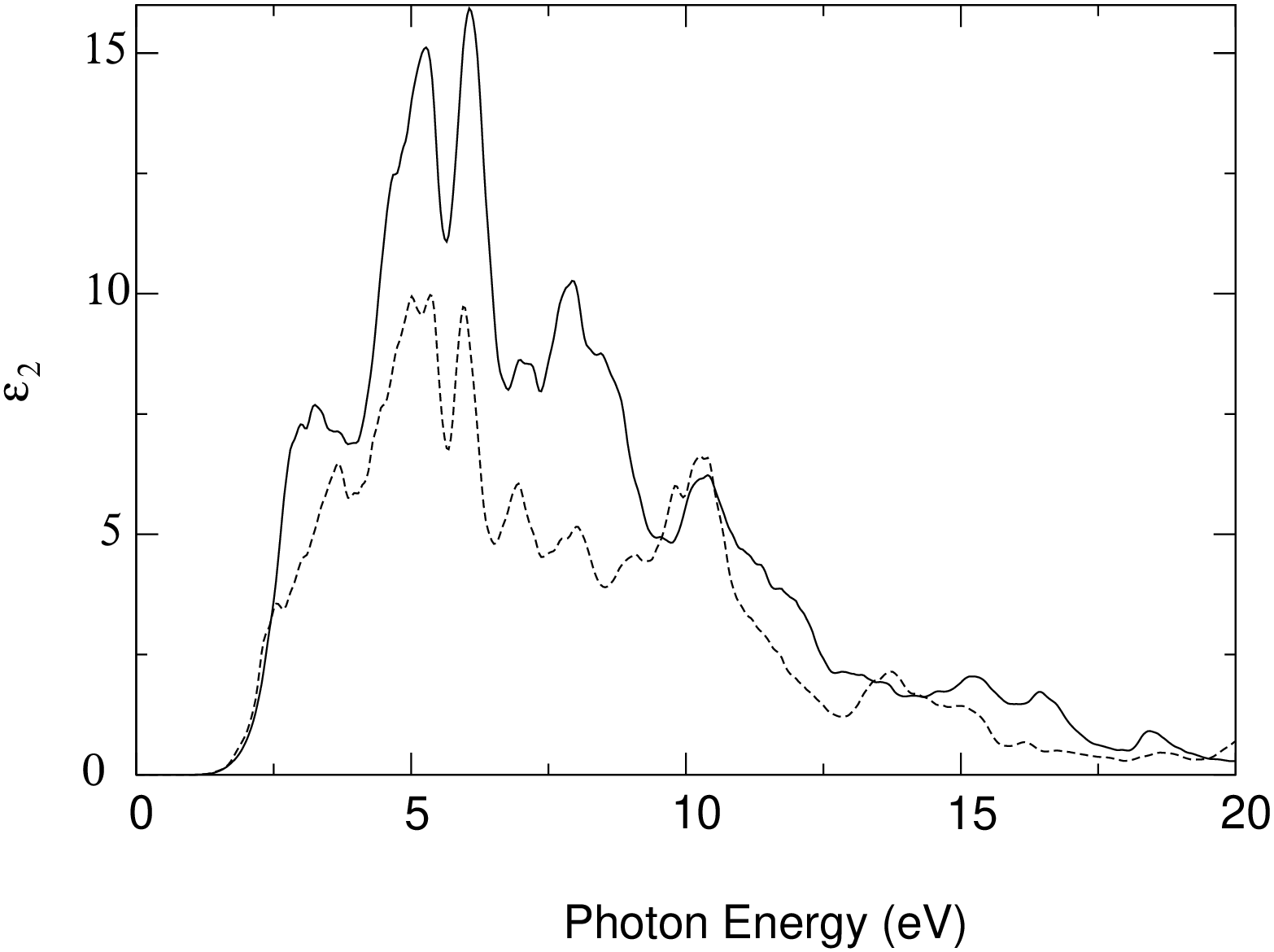}}
\rotatebox{-90}{\epsfbox{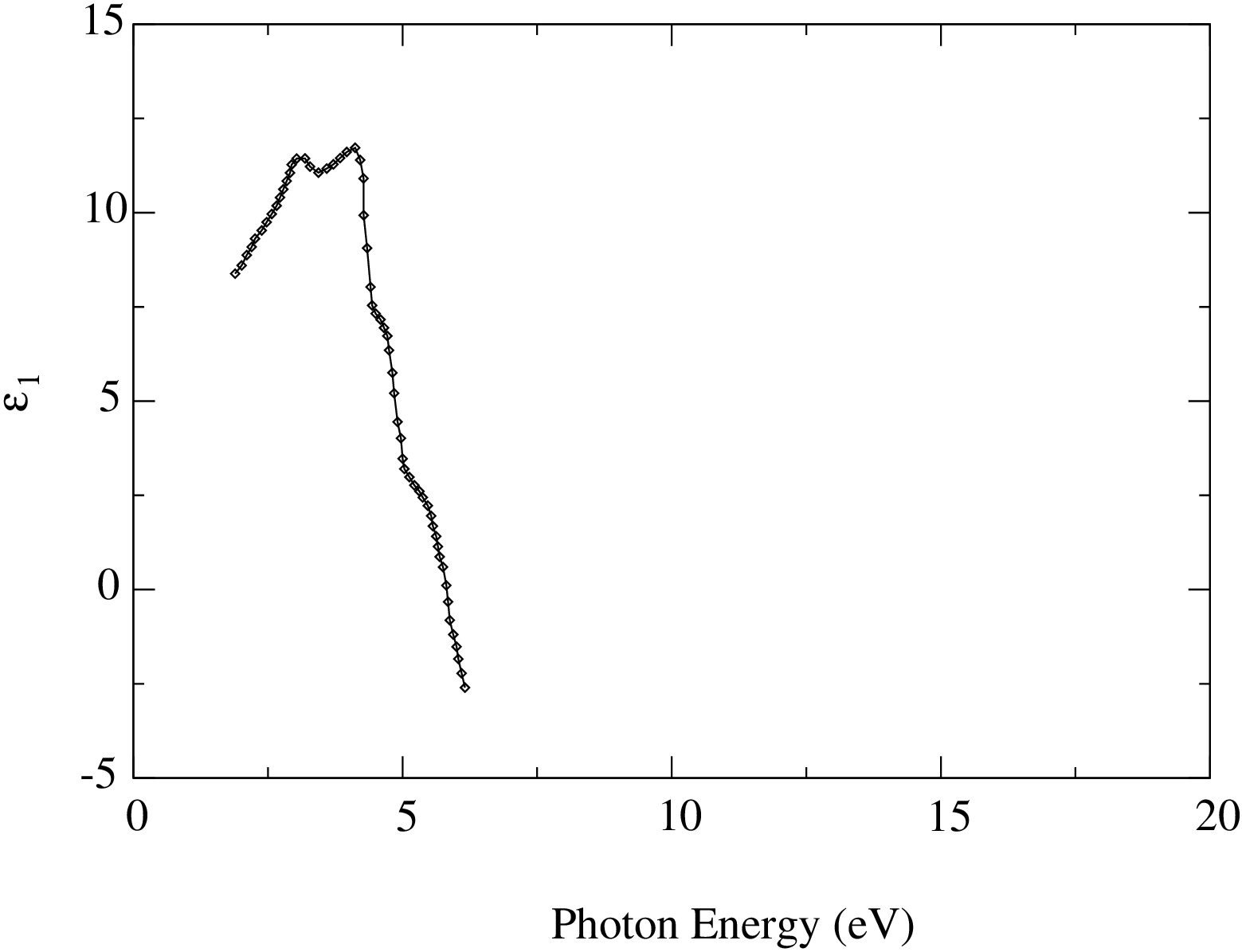}}
\rotatebox{-90}{\epsfbox{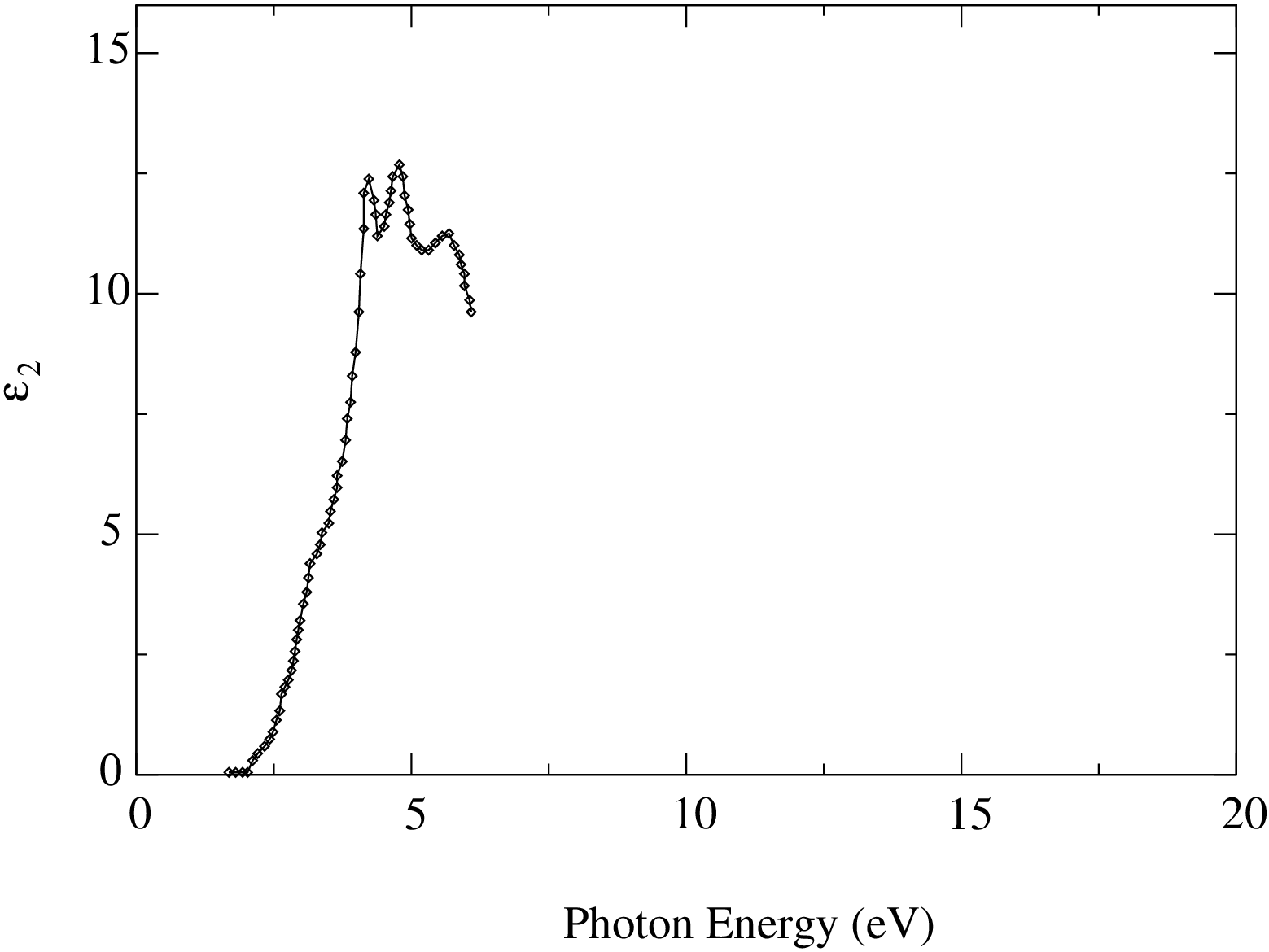}}
\caption{(Top) Real and Imaginary parts of the dielectric function of $ZnIn_2Te_4$ calculated from
an NMTO calculation. The full lines show $\epsilon_{\perp}$ and the dashed lines $\epsilon_{\parallel}$.
(Bottom) The experimental results of Ozaki \etal \cite{Ozaki}}
\label{opt}
\end{figure}

The lattice for $ZnIn_2Te_4$ does not have cubic symmetry. If we take the $c$-axis to be the z-axis, it
is clear that the three optical responses $\epsilon_x$, $\epsilon_y$ and $\epsilon_z$ are not the same.
In the top part of figure \ref{opt}, we show the variation of  the real and imaginary parts of
$\epsilon_{\parallel}$  = $\epsilon_z$ and $\epsilon_\perp$ = $\epsilon_x+\epsilon_y$ with the incident photon energy 

We first note that since our LDA-based calculations give a smaller band gap that experiment, we have
applied the {\it scissors} operator, which involves a rigid shift of the conduction band with respect
to the valence band, so that the band gap matches. This is frequently used by LDA practitioners, but
cannot be fully justified. The correct procedure would be to carry out a full GW calculation which gives
rise to an {\sl energy dependent} self-energy, which shifts the bands unequally at different
energies, resulting in a distortion of the shape of the densities of states as well. Given this, the
agreement of the theoretical calculations with the experimental results of Ozaki \etal \cite{Ozaki}
available up to 10 eV photon energies, is not bad.  The experiment does not align the crystal axis
with the polarization of the electric field, so that we should compare the results the direction
averaged response $(\epsilon_\parallel +\epsilon_\perp)/3$. The principal peak positions and heights are well
reproduced.

We shall conclude with the remark on the two directions which we propose to take from here.
The first point is to recognize that the NMTO calculations form a reasonable starting
point of the more sophisticated many-body GW approaches. The second is related to the reason why we chose
to study the defect chalcopyrite in the first place. We would like to fill the voids in the structure with
various ``impurities" and study the signal for these in the optical response. 
These will be the aim of our subsequent work in this area.

\ack We would like to thank Prof. O.K. Andersen for kind permission to use the NMTO codes developed by his
group. Financial help from the University of Warwick is also gratefully acknowledged.

\end{document}